\documentclass{elsart}
\usepackage{setspace,times,amssymb,amsmath,multirow,graphicx,color,rotating,subfigure,url}
\usepackage{lineno}
\usepackage[square,sort&compress,comma]{natbib}

\journal{Physica A} 

\begin{document}

\begin{frontmatter}

\title{On the probability distribution of stock returns in the Mike-Farmer model}
\author[BS,SS]{Gao-Feng Gu},
\author[BS,SS,RCE,RCSE]{Wei-Xing Zhou \corauthref{cor}}
\corauth[cor]{Corresponding author. Address: 130 Meilong Road, P.O.
Box 114, School of Business, East China University of Science and
Technology, Shanghai 200237, China, Phone: +86 21 64253634, Fax: +86
21 64253152.}
\ead{wxzhou@ecust.edu.cn} %

\address[BS]{School of Business, East China University of Science and Technology, Shanghai 200237, China}
\address[SS]{School of Science, East China University of Science and Technology, Shanghai 200237, China}
\address[RCE]{Research Center for Econophysics, East China University of Science and Technology, Shanghai 200237, China}
\address[RCSE]{Research Center of Systems Engineering, East China University of Science and Technology, Shanghai 200237, China}

\begin{abstract}
Recently, Mike and Farmer have constructed a very powerful and
realistic behavioral model to mimick the dynamic process of stock
price formation based on the empirical regularities of order
placement and cancelation in a purely order-driven market, which can
successfully reproduce the whole distribution of returns, not only
the well-known power-law tails, together with several other
important stylized facts. There are three key ingredients in the
Mike-Farmer (MF) model: the long memory of order signs characterized
by the Hurst index $H_s$, the distribution of relative order prices
$x$ in reference to the same best price described by a Student
distribution (or Tsallis' $q$-Gaussian), and the dynamics of order
cancelation. They showed that different values of the Hurst index
$H_s$ and the freedom degree $\alpha_x$ of the Student distribution
can always produce power-law tails in the return distribution $f(r)$
with different tail exponent $\alpha_r$. In this paper, we study the
origin of the power-law tails of the return distribution $f(r)$ in
the MF model, based on extensive simulations with different
combinations of the left part $f_L(x)$ for $x<0$ and the right part
$f_R(x)$ for $x>0$ of $f(x)$. We find that power-law tails appear
only when $f_L(x)$ has a power-law tail, no matter $f_R(x)$ has a
power-law tail or not. In addition, we find that the distributions
of returns in the MF model at different timescales can be well
modeled by the Student distributions, whose tail exponents are close
to the well-known cubic law and increase with the timescale.
\end{abstract}

\end{frontmatter}

\section{Introduction}
\label{S1:MFM:Intro}

Many stylized facts have been unveiled in different stock markets
\cite{Mantegna-Stanley-2000,Cont-2001-QF,Zhou-2007}. Understanding
the underlying regularities causing stylized facts are crucial in
stock market modeling. Three different families of market models
exist aiming at reproducing the main stylized facts. The first
family is the dynamic models, such as the multifractal model of
asset returns \cite{Mandelbrot-Fisher-Calvet-1997}, which was later
extended in several publications
\cite{Calvet-Fisher-2001-JEm,Lux-2003,Lux-2004,Eisler-Kertesz-2004-PA},
and the multifractal random walks
\cite{Bacry-Delour-Muzy-2001-PA,Bacry-Delour-Muzy-2001-PRE,Pochart-Bouchaud-2002-QF}.
The second family is the agent-based models (or multi-agent models),
in which agents buy or sell shares according to some rules and the
price variations are determined by the imbalance of demand and
supply. There are different types of agent-based models, such as
percolation models
\cite{Cont-Bouchaud-2000-MeD,Stauffer-1998-AP,Stauffer-Penna-1998-PA,Eguiluz-Zimmermann-2000-PRL,DHulst-Rodgers-2000-IJTAF,Xie-Wang-Quan-Yang-Hui-2002-PRE},
Ising models
\cite{Foellmer-1974-JMe,Chowdhury-Stauffer-1999-EPJB,Iori-1999-IJMPC,Kaizoji-2000-PA,Bornholdt-2001-IJMPC,Zhou-Sornette-2007-EPJB},
minority games
\cite{Arthur-1994-AER,Challet-Zhang-1997-PA,Challet-Marsili-Zhang-2000-PA,Jefferies-Hart-Hui-Johnson-2001-EPJB,Challet-Marsili-Zhang-2001-QF,Challet-Marsili-Zhang-2001a-PA,Challet-Marsili-Zhang-2001b-PA,Challet-Marsili-Zhang-2005},
and others \cite{Lux-Marchesi-1999-Nature}. The minority games are
among the most important agent-based models and thus many variants
have been proposed. The third family is the order-driven models,
where researchers attempt to simulate the dynamics of order books.
The price in order-driven models changes based on the continuous
double auction (CDA) mechanism
\cite{Maslov-2000-PA,Daniels-Farmer-Gillemot-Iori-Smith-2003-PRL,Farmer-Patelli-Zovko-2005-PNAS}.
A nice review of order-driven models was recently given by Slanina
\cite{Slanina-2008-EPJB}. We can think of the agent-based models and
the order-driven models as microscopic models for quote-driven
markets and order-driven markets, respectively.

Recently, Mike and Farmer have constructed a very powerful and
realistic behavioral model to mimick the dynamic process of stock
price formation \cite{Mike-Farmer-2008-JEDC}, which belongs to the
third family. We call it {\em{Mike-Farmer model}}, or MF model for
short. It seems undoubtable to us that the MF model is a milestone
in the modeling of order-driven markets, which will prove to
introduce an important improvement in asset derivative pricing and
risk management. Having said this, we stress that the MF model is
still very simple as mentioned already by Mike and Farmer
\cite{Mike-Farmer-2008-JEDC} and there are still a lot of open
problems to be addressed. Indeed, the MF model provides a nice
platform to unravel the origin of stylized facts of stocks in
order-driven markets. The essential advantage of the MF model is
that it is constructed based on the empirical regularities of order
placement and cancelation in a purely order-driven market, which can
successfully reproduce the whole distribution of returns, not only
the well-known power-law tails, together with several other
important stylized facts. There are three key ingredients in the MF
model: the long memory of order signs characterized by the Hurst
index $H_s$, the distribution of relative order prices $x$ in
reference to the same best price described by a Student
distribution, and the dynamics of order cancelation.

Through extensive simulations, Mike and Farmer found that different
values of the Hurst index $H_s$ and the freedom degree $\alpha_x$ of
the Student distribution can always produce power-law tails in the
return distribution $f(r)$ with different tail exponent $\alpha_r$.
Specifically, they found that $\alpha_r$ increases almost linearly
with $\alpha_x$ for fixed $H_s$ and decreases approximately linearly
with $H_s$ for fixed $\alpha_x$. Our simulations of the MF model
with different values of $\alpha_x$ (ranging from 0.9 to 1.9 with a
step of 0.1) and $H_s$ (ranging from 0.1 to 0.9 with a step of 0.1)
confirm this finding. Speaking differently we have simulated
$9\times11$ versions of the MF model. For each version, four million
simulation steps are conducted after removing the initial transient
data and we obtain about one million data points of trade-by-trade
returns. We find that the dependence of $\alpha_r$ upon $H_s$ and
$\alpha_x$ can be modeled using the following formula:
\begin{equation}
 \alpha_r = 0.61+2.05\alpha_x-0.11 H_s-0.34 H_s\alpha_x~.
 \label{Eq:MFM:alphar:alphax:H}
\end{equation}
These observations can be explained as follows. With the increase of
$H_s$, the memory of order signs becomes stronger and more orders of
the same direction (buy or sell) are placed successively. This
results in more large price fluctuations and the decay of $f(r)$
becomes slower. Hence the tail exponent $\alpha_r$ decreases. On the
other hand, if $\alpha_x$ is small, more passive orders with $x<0$
are placed deep inside the order book and the standing volumes close
to the best ask or bid price are relatively small. Speaking
differently, the depth of the order book is low and the liquidity is
low. Also, there are more aggressive orders with $x>0$ placed
resulting in more market orders. Both effects lead to more large
price fluctuations and slower decay of the return distribution.

Although both the strength of the long memory of order signs and the
tail exponent of relative order prices have significant influence on
the distribution of returns, it is unclear what causes the power-law
tails in the MF model. In this paper, we will address this question
based on extensive simulations with different combinations of the
left part $f_L(x)$ for $x<0$ and the right part $f_R(x)$ for $x>0$
of $f(x)$. We find that power-law tails appear only when $f_L(x)$
has a power-law tail, no matter $f_R(x)$ has a power-law tail or
not. Moreover, we find that the return distributions in the MF model
at different timescales can be well modeled by the Student (or
Tsallis' $q$-Gaussian) distributions, whose tail exponents are close
to the well-known cubic law and increase with the timescale.

\section{Description of the Mike-Farmer model}
\label{S1:MFM:ModelDescription}

In purely order-driven markets, the main trading mechanism is the
continuous double auction. Passive traders are patient and place
effective limit orders that are stored in the order book waiting for
execution, while aggressive traders are inpatient and submit
effective market orders that are executed immediately. Consider a
limit order placed at event time $t$ whose logarithmic price is
$\pi(t)$. Denote $\pi_a(t-1)$ and $\pi_b(t-1)$ the logarithms of
best ask and bid prices right before $t$. A buy limit order with
$\pi(t)\geqslant\pi_a(t-1)$ or a sell limit order with
$\pi(t)\leqslant\pi_b(t-1)$ is classified as an effective market
order. Speaking differently, orders with the relative prices $x(t)$
less than the preceding spreads $S(t-1)=\pi_a(t-1)-\pi_b(t-1)$ are
effective limit orders, while orders with $x(t)\geqslant S(t-1)$ are
effective market orders\footnote{The relative price $x(t)$ is
defined as $\pi(t)-\pi_b(t-1)$ for buy orders and
$\pi_a(t-1)-\pi(t)$ for sell orders.}. Orders waiting on the limit
order book are either satisfied by future effective market orders or
canceled. Therefore, the continuous double auction can be simulated
if one knows the regularities governing the dynamic processes of
order placement and cancelation. Most order-driven models also
follow this line. However, to the best of our knowledge, the MF
model is the only one that uses empirical regularities of order
placement and cancelation extracted from real stock data. This is
the reason why the MF model can reproduce the cubic law of return
distribution without tuning any model parameters. Actually, the MF
model does not introduce any artificial tunable model parameters at
all. All parameters in the MF model are determined empirically and
have clear financial meanings. The regularities of order placement
and cancelation may be different for different stock markets. The MF
model can be easily modified for other markets.

When placing an order, the trader needs to determine its sign
(``$+1$'' for buys and ``$-1$'' for sells), size and price (or the
relative price $x$). In the MF model, all orders are assumed to have
identical size. The signs of successive orders have strong memory,
which can be characterized by a rather large Hurst index $H_s$ close
to 0.8 or even larger
\cite{Lillo-Farmer-2004-SNDE,Mike-Farmer-2008-JEDC}. This finding is
conclusive without any controversy. In contrast, the distribution of
relative prices seems different in different markets. Zovko and
Farmer studied the unconditional distribution of relative limit
prices defined as the distance from the same best prices for orders
placed inside the limit-order book \cite{Zovko-Farmer-2002-QF}. They
merged the data from 50 stocks traded on the London Stock Exchange
(August 1, 1998 to April 31, 2000) and found that the distribution
decays roughly as a power law with the tail exponent $\alpha = 1.5$
for both buy and sell orders. Bouchaud {\em et al}. analyzed the
order books of three liquid stocks on the Paris Bourse (February
2001) and found that the relative price of new orders placed inside
the book follows a power-law distribution with the tail exponent
$\alpha = 0.6$ \cite{Bouchaud-Mezard-Potters-2002-QF}. Potters and
Bouchaud investigated the relative limit price distributions for
inside-the-book orders of three Nasdaq stocks (June 1 to July 15,
2002) and found that the distributions exhibit power-law tails with
an exponent $\alpha = 1$ \cite{Potters-Bouchaud-2003-PA}. Maskawa
analyzed 13 rebuild order books of Stock Exchange Electronic Trading
Service from July to December in 2004 on the London Stock Exchange
and found that the limit prices for all orders inside the book are
broadly distributed with a power-law tail whose exponent is $\alpha
= 1.5$ \cite{Maskawa-2007-PA}, which is consistent with the results
of Zovko and Farmer \cite{Zovko-Farmer-2002-QF}. He also presented
the distribution in the negative part for more aggressive order
outside the book and found that the negative part decays much faster
than the positive part. Mike and Farmer focused on the stock named
AZN and tested on 24 other stocks listed on the London Stock
Exchange (LSE) \cite{Mike-Farmer-2008-JEDC}. They found that the
distribution of relative logarithmic prices can be fitted by a
Student distribution with $\alpha=1.0-1.65$ degrees of freedom and
the distribution is independent of bid-ask spread at least over a
restricted range for both buy and sell orders. Gu, Chen and Zhou
analyzed 23 Chinese stocks traded on the Shenzhen Stock Exchange
(SZSE) and found that the distribution of relative prices is
asymmetric \cite{Gu-Chen-Zhou-2008b-PA}. They showed that the
distribution has power laws with the exponents greater than 1 and
lower than 2.

There are also efforts to seek for factors influencing order
placement. Using 15 stocks on the Swiss Stock Exchange, Ranaldo
found that both bid-ask spread and volatility negatively relate to
order aggressiveness \cite{Ranaldo-2004-JFinM}. Lillo analyzed the
origin of power-law distribution of limit order prices considering
the order placement as an utility maximization problem considering
three factors: time horizon, utility function and volatility
\cite{Lillo-2007-EPJB}. He found that the heterogeneity in time
horizon is the proximate cause of the asymptotic power-law
distribution, while heterogeneity in volatility is hardly connected
with the origin of power-law distribution. Mike and Farmer found
that the distribution $f(x)$ of LSE stocks is independent of the
bid-ask spread \cite{Mike-Farmer-2008-JEDC}, which was confirmed by
Gu, Chen and Zhou using SZSE stocks \cite{Gu-Chen-Zhou-2008b-PA}.

In a zero intelligence model
\cite{Daniels-Farmer-Gillemot-Iori-Smith-2003-PRL,Farmer-Patelli-Zovko-2005-PNAS},
order cancelation is assumed to be a Poisson process. Alternatively,
Mike and Farmer found that the conditional probability of canceling
an order $i$ at time $t$ is influenced by at least three factors
\cite{Mike-Farmer-2008-JEDC}: the ratio $y_i$ of current relative
price of an order to its original relative price when it is placed,
the number $n_{\rm{tot}}$ of orders in the order book, and the order
book imbalance $n_{\rm{imb}}$ that is defined as the ratio of the
number of buy (or sell) orders to $n_{\rm{tot}}$. By assuming that
$y_i$, $n_{\rm{tot}}$ and $n_{\rm{imb}}$ are independent, the
conditional probability of cancelation per order has the following
form:
\begin{equation}
 P(C_i|y_i, n_{\rm{tot}}, n_{\rm{imb}}) =
 A(1-e^{-y_i})(n_{\rm{imb}}+B)/n_{\rm{tot}}~,
 \label{Eq:P:Ci}
\end{equation}
where the parameters $A$ and $B$ are determined empirically using
real data of individual stocks \cite{Mike-Farmer-2008-JEDC}.

Now we can describe the MF model as follows. We stress that the
sizes of orders are set to unity. In each round of the simulation,
we simulate $2\times10^5$ steps and the first 2000 data points are
discarded from the ensuing analysis. We repeat this process 20
times, which results in four million orders and about one million
transactions. In each round, we generate two arrays of the relative
prices $x(t)$ and the order signs $s(t)$. The sign array of the
orders is generated from a fractional Brownian motion with Hurst
index $H_s$ and the relative price $x$ is taken from a Student
distribution with scale $\sigma_x$ and $\alpha_x$ degrees of
freedom. At each simulation step or event time $t$, an order is
generated, which is characterized by $x(t)$ and $s(t)$. When
$x(t)\geqslant S(t-1)$, the order is executed and a buy limit order
(if $s(t)<0$) or a sell limit order (if $s(t)>0$) at the best bid or
ask price is removed from the order book. When $x(t)< S(t-1)$, the
order is stored in the order book at the price level
$T{\rm{int}}[{\pi(t)}/T]$, where $T$ is the tick size,
${\rm{int}}[z]$ is the largest integer smaller than $z$, and
$\pi(t)= x(t)+\pi_b(t-1)$ for buy orders or $\pi(t)=
\pi_a(t-1)-x(t)$ for sell orders. We then calculate the values of
$P(C_i|y_i, n_{\rm{tot}}, n_{\rm{imb}})$ for all orders waiting in
the order book. A random number $p(t)$ is drawn from a uniform
distribution defined on the interval $[0,1]$. All orders with
$P(C_i|y_i, n_{\rm{tot}}, n_{\rm{imb}})\leqslant p(t)$ are canceled
from the order book. The returns between successive trades are used
in this work. We note that the results presented in
\cite{Mike-Farmer-2008-JEDC} are successfully reproduced in our
simulations.

\section{The origin of power-law tail of returns}
\label{S1:MFM:WhyPL}

\subsection{Methodology}
\label{S2:MFM:Methodology}

Mike and Farmer have shown that the power-law tails in the return
distribution $f(r)$ become heavier if the long memory in the order
signs is stronger \cite{Mike-Farmer-2008-JEDC}. We notice that the
power-law tails do not vanish even when $H_s=0.5$. In other words,
long memory in the order signs cannot explain the emergence of
power-law tails of returns. Therefore, we turn to investigate the
influence of $f(x)$ on $f(r)$. According to the setting of the MF
model, the shape of $f(x)$ for $x>S$ does not impact the shape of
$f(r)$, since those orders are effective market orders that always
remove one unit of shares from the opposite side of the order book,
despite of the true prices of the incoming effective market orders.
We thus speculate that the right part of $f(x)$, denoted as
$f_R(x)$, has much weaker influence than the left part $f_L(x)$.  In
our simulations, the tick size is $T=3\times10^{-4}$ and three model
parameters are fixed according to \cite{Mike-Farmer-2008-JEDC}:
$H_s=0.8$, $A=1.12$, and $B=0.2$.

In order to unveil the effect of the two parts of $f(x)$ on the tail
behavior of $f(r)$, we adopt different formulae for $f(x)$. In
general, we can write the following
\begin{equation}
 f(x) = \left\{
 \begin{array}{ccc}
 f_L(x),&& x\leqslant0\\
 f_R(x),&& x\geqslant0
 \end{array}
 \right.~.
 \label{Eq:fx:LR}
\end{equation}
Obviously, when $x=0$, we require that
\begin{equation}
 f_L(0)= f_R(0)~.
 \label{Eq:fx:LR:0}
\end{equation}
In the MF model, $f(x)=f_L(x)= f_R(x)$. In this section, we will use
different functional forms for $f_L(x)$ and $f_R(x)$.

The first class is the Student density
\cite{Blattberg-Gonedes-1974-JB} or Tsallis' $q$-Gaussian
\cite{Tsallis-Anteneodo-Borland-Osorio-2003-PA}, whose density is
\begin{equation}
f_{qG}(x|\alpha_x,L) =
\frac{\sqrt{L}\alpha_x^{\frac{\alpha_x}{2}}}{B\left(\frac{1}{2},\frac{\alpha_x}{2}\right)}\left(\alpha_x
+ Lx^2\right)^{-\frac{\alpha_x + 1}{2}}~,
 \label{Eq:Student}
\end{equation}
where $\alpha_x$ is the degrees of freedom parameter (or tail
exponent), $L$ is the scale parameter, and $B(a, b)$ is the Beta
function, that is, $B(a, ~b) = \Gamma(a)~\Gamma(b) ~/ ~\Gamma(a +
b)$ with $\Gamma(\cdot)$ being the gamma function. The second class
is the Laplace distribution or double exponential distribution,
whose density is
\begin{equation}
f_{DE}(x|\lambda) = \frac{\lambda}{2}e^{-\lambda|x|}~,
 \label{Eq:Laplace}
\end{equation}
and the third class is the normal distribution, whose density is
\begin{equation}
f_{G}(x|\sigma) = \frac{1}{\sqrt{2\pi}\sigma}e^{-x^2/2\sigma^2}~,
 \label{Eq:Gaussian}
\end{equation}
According to the symmetry of the distribution of $x$ in the MF model
\cite{Mike-Farmer-2008-JEDC}, the mean of $x$ is fixed to null in
all these distributions.

In the model specification, we choose $f_L(x)$ and $f_R(x)$ from
$f_{qG}$, $f_{DE}$ and $f_G$, respectively. The constraint
(\ref{Eq:fx:LR:0}) can be specified for different combination of
$f_L(x)$ and $f_R(x)$ as follows. When the two parts of $f(x)$ are
$f_{qG}$ and $f_{DE}$, we have
\begin{equation}
 2\sqrt{{L}/{\alpha_x}}
 = \lambda B\left({1}/{2},{\alpha_x}/{2}\right)~.
 \label{Eq:x0:qG:DE}
\end{equation}
When the two parts of $f(x)$ are $f_{qG}$ and $f_{G}$, we have
\begin{equation}
 B\left({1}/{2},{\alpha_x}/{2}\right) = \sigma\sqrt{{2\pi L}/{\alpha_x}}~.
 \label{Eq:x0:qG:G}
\end{equation}
When the two parts of $f(x)$ are $f_{DE}$ and $f_{G}$, we have
\begin{equation}
 \lambda^2\sigma^2 = {2/\pi}~,
 \label{Eq:x0:DE:G}
\end{equation}
which is equivalent to the combination of Eq.~(\ref{Eq:x0:qG:DE})
and Eq.~(\ref{Eq:x0:qG:G}).

We fix $\sigma_x = 0.0024$ in all the simulations so that the
simulated sample of $x$ is comparable to real data
\cite{Mike-Farmer-2008-JEDC}. The values of $L$ in
Eq.~(\ref{Eq:Student}) are calculated as follows:
\begin{equation}
 L = \alpha_x/[(1+\alpha_x)\sigma_x^2]~.
 \label{Eq:L:alphax}
\end{equation}
The values of $\lambda$ and $\sigma$ are determined respectively
according to Eq.~(\ref{Eq:x0:qG:DE}) and Eq.~(\ref{Eq:x0:qG:G}) for
different values of $\alpha_x$. We investigate different
combinations of $f_L(x)$ and $f_R(x)$ according to if they have a
power-law tail. The case that both parts have a power-law tail has
been studied by Mike and Farmer \cite{Mike-Farmer-2008-JEDC}, as
discussed in Section \ref{S1:MFM:Intro}. Therefore, we are left with
three cases: (1) There are no power-law tails in $f(x)$; (2) The
right part of $f(x)$ has a power-law tail with
$f_R(x)=f_{qG}(x|\alpha_x,L)$; and (3) The left part of $f(x)$ has a
power-law tail with $f_L(x)=f_{qG}(x|\alpha_x,L)$. We shall
investigate these three cases in the rest of this section.

\subsection{Case 1: There are no power-law tails in $f(x)$}
\label{S2:MFM:Case1}

In this case, there are no power-law tails in $f(x)$ and we have
four combinations for $f_L(x)$ and $f_R(x)$ that are
$\{f_{DE}(x|\lambda), f_G(x|\sigma)\}$, $\{f_{DE}(x|\lambda),
f_{DE}(x|\lambda)\}$, $\{f_{G}(x|\sigma), f_{DE}(x|\lambda)\}$, and
$\{f_{G}(x|\sigma), f_G(x|\sigma)\}$. For each combination, we
investigate five different values of $\alpha_x$ from 1.1 to 1.9 with
a step of 0.2. For each value of $\alpha_x$, our simulations are
conducted for 20 repeated rounds. In each round, $2\times10^5$
incoming orders are generated, driving the trading system evolve,
and the first 2000 orders are excluded from analysis. The
complementary cumulative distribution can be determined for each
$\alpha_x$ in every combination. We show the tails of $F(|r|)$,
which is the complementary cumulative probability distribution of
$|r|$, since the return distributions are almost symmetric for
positive and negative returns.

\begin{figure}[htb]
\centering
\includegraphics[width=6.5cm]{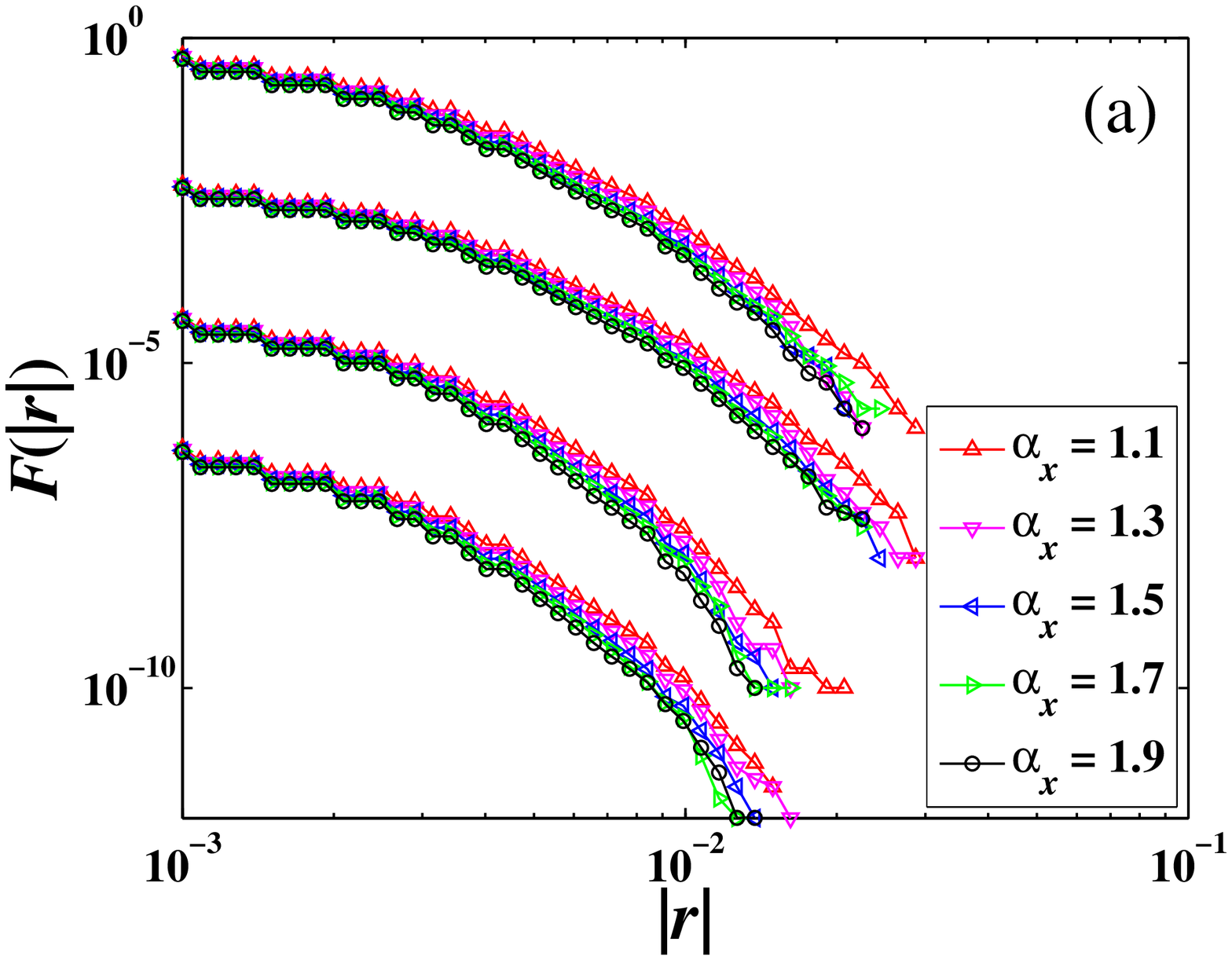}
\includegraphics[width=6.5cm]{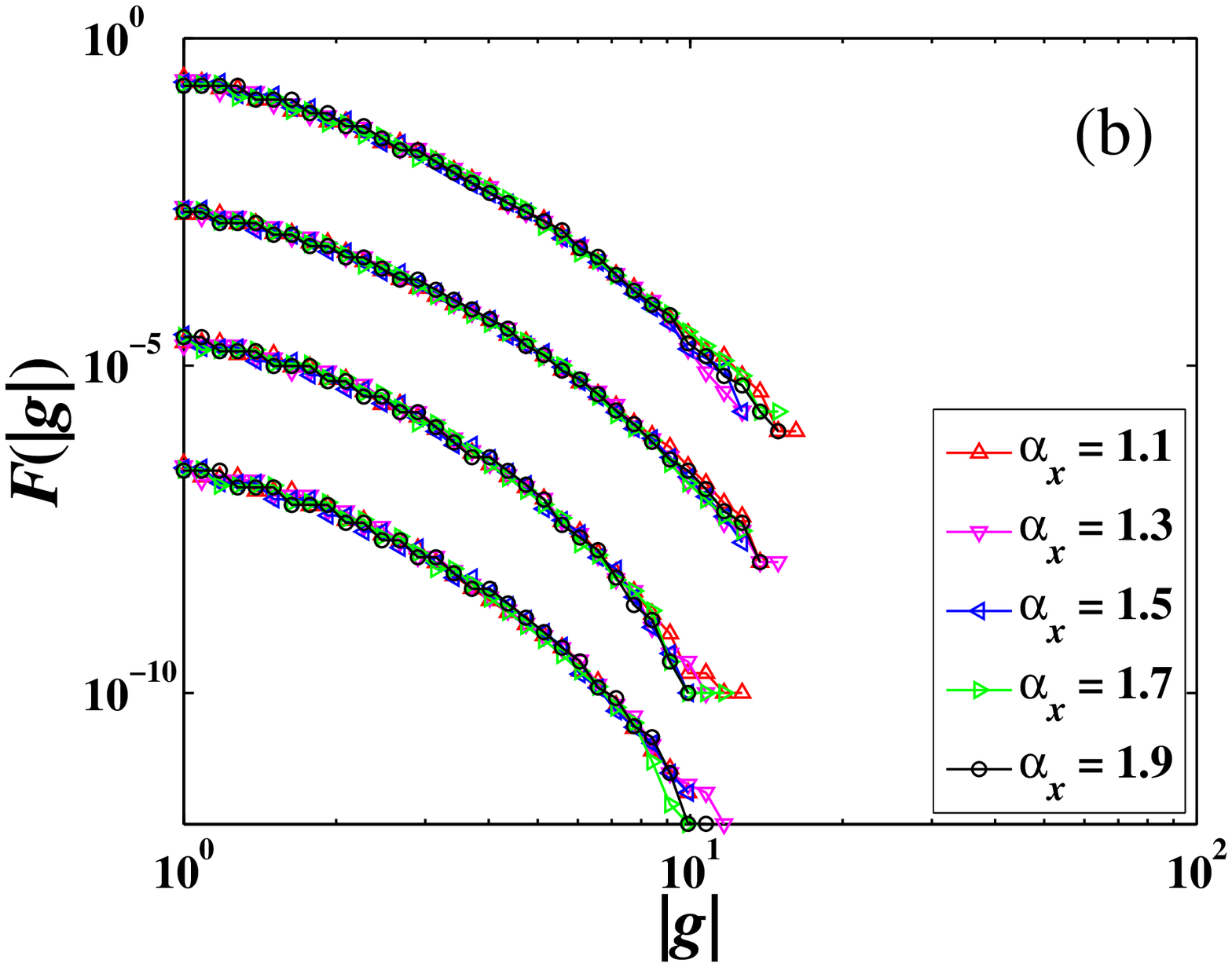}
\caption{\label{Fig:CDF:DE:G} (Color online) Resultant distribution
of the trade-by-trade absolute returns $|r|$ (a) and the
corresponding standardized returns $|g|$ (b) when the relative price
distribution $f(x)$ has no power-law tails. Each cluster of curves
corresponds to a combination of $f_L(x)$ and $f_R(x)$. The clusters
have been shifted vertically for clarification. The associated
combination $\{f_L(x),f_R(x)\}$ from top to bottom is
$\{f_{DE}(x|\lambda),f_G(x|\sigma)\}$, $\{f_{DE}(x|\lambda),
f_{DE}(x|\lambda)\}$, $\{f_{G}(x|\sigma), f_{DE}(x|\lambda)\}$, and
$\{f_{G}(x|\sigma), f_G(x|\sigma)\}$, where $\lambda$ and $\sigma$
are determined respectively according to Eq.~(\ref{Eq:x0:qG:DE}) and
Eq.~(\ref{Eq:x0:qG:G}) for different values of $\alpha_x$.}
\end{figure}

The resultant 20 empirical distributions of the trade-by-trade
absolute returns $|r|$ for each combination are given in
Fig.~\ref{Fig:CDF:DE:G}(a). It is evident that there is no power-law
tails observed in the distributions. For each combination, the
distribution $F(|r|)$ decays faster for larger $\alpha_x$. We also
find that there are more large returns for the two combinations
where $f_L=\{f_{DE}(x|\lambda)$. This is explained by the two facts
that $f_L(x)$ has stronger impact on $F(|r|)$ or equivalently $f(r)$
and that a Laplace distribution has heavier tails than a Gaussian.
Fig.~\ref{Fig:CDF:DE:G}(b) presents the distributions of the
standardized returns $g=(r-\mu_r)/\sigma_r$, where $\mu_r(\approx0)$
and $\sigma_r$ are respectively the mean and the standard deviation
of $r$. It is interesting to observe that the five distributions
collapse onto a single curve for all four combinations.

\subsection{Case 2: The right part of $f(x)$ has a power-law tail with $f_R(x)=f_{qG}(x|\alpha_x,L)$}
\label{S2:MFM:Case2}

\begin{figure}[htb]
\centering
\includegraphics[width=6.5cm]{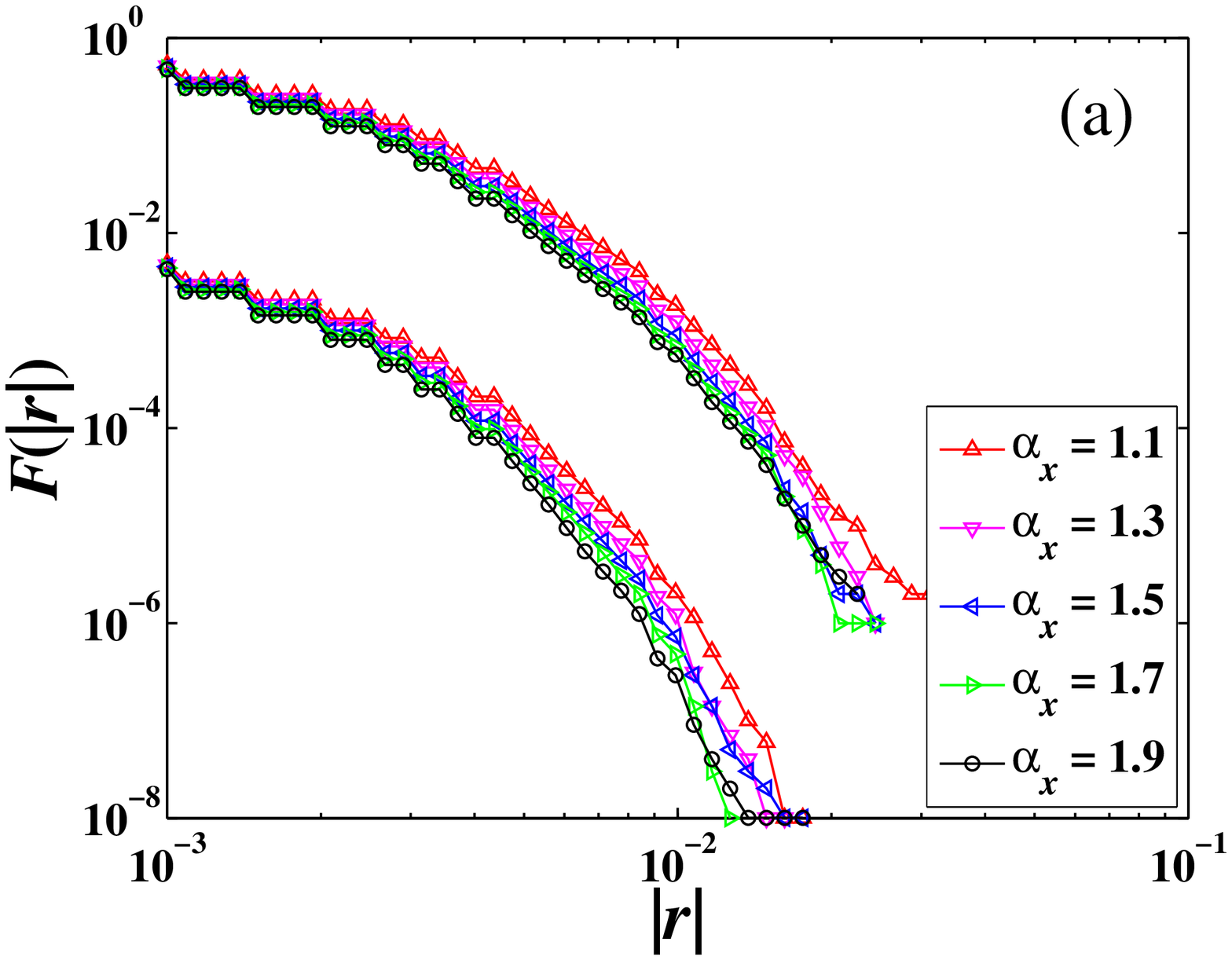}
\includegraphics[width=6.5cm]{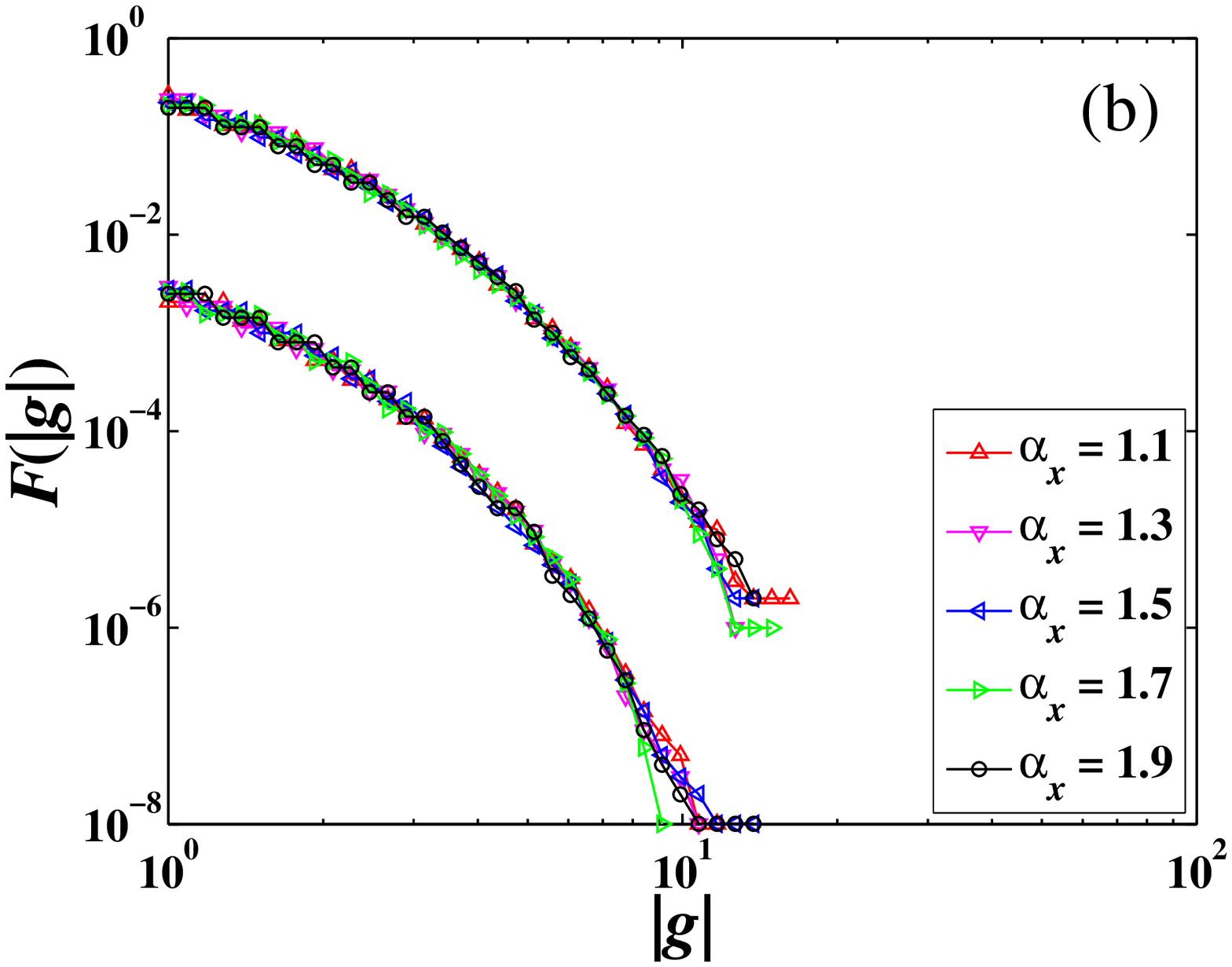}
\caption{\label{Fig:CDF:G:DE:qG} (Color online) Resultant
distribution of the trade-by-trade absolute returns $|r|$ (a) and
the corresponding normalized returns $|g|$ (b) when the right part
of $f(x)$ has a power-law tail, that is,
$f_R(x)=f_{qG}(x|\alpha_x,L)$. The upper cluster of curves
corresponds to $f_L(x)=f_{DE}(x|\lambda)$ and the lower cluster
corresponds to $f_L(x)=f_{G}(x|\sigma)$, where $\lambda$ and
$\sigma$ are determined respectively according to
Eq.~(\ref{Eq:x0:qG:DE}) and Eq.~(\ref{Eq:x0:qG:G}) for different
values of $\alpha_x$. The lower cluster has been shifted downwards
for clarification.}
\end{figure}

In this case, the right part of $f(x)$ has a power-law tail with
$f_R(x)=f_{qG}(x|\alpha_x,L)$ and we have two combinations for
$f_L(x)$ and $f_R(x)$: $\{f_{DE}(x|\lambda),f_{qG}(x|\alpha_x,L)\}$
and $\{f_{G}(x|\sigma), f_{qG}(x|\alpha_x,L)\}$. The simulation
procedure is the same as in Section \ref{S2:MFM:Case1}. The
resultant 10 empirical distributions of the trade-by-trade absolute
returns $|r|$ for each combination are given in
Fig.~\ref{Fig:CDF:G:DE:qG}(a). Again, no power-law tails are
observed in these distributions. For each combination, the
distribution $F(|r|)$ decays faster for larger $\alpha_x$. We also
find that there are more large returns for the the combination where
$f_L=\{f_{DE}(x|\lambda)\}$. The same explanation applies.
Fig.~\ref{Fig:CDF:G:DE:qG}(b) presents the distributions of the
normalized returns $|g|$ and the five distributions for each
combination collapse remarkably onto a single curve.

\subsection{Case 3: The left part of $f(x)$ has a power-law tail with $f_L(x)=f_{qG}(x|\alpha_x,L)$}
\label{S2:MFM:Case3}

\begin{figure}[htb]
\centering
\includegraphics[width=6.5cm]{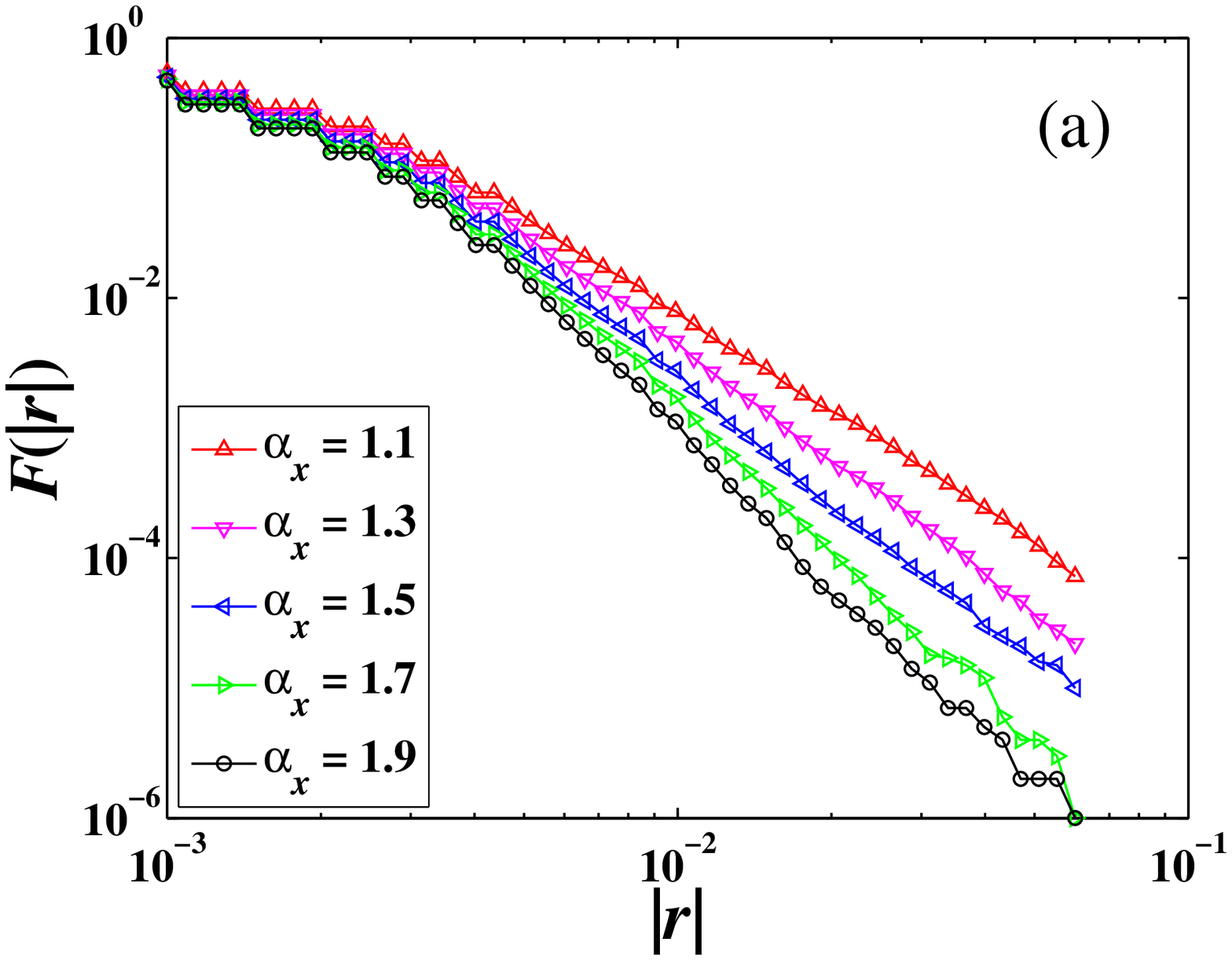}
\includegraphics[width=6.5cm]{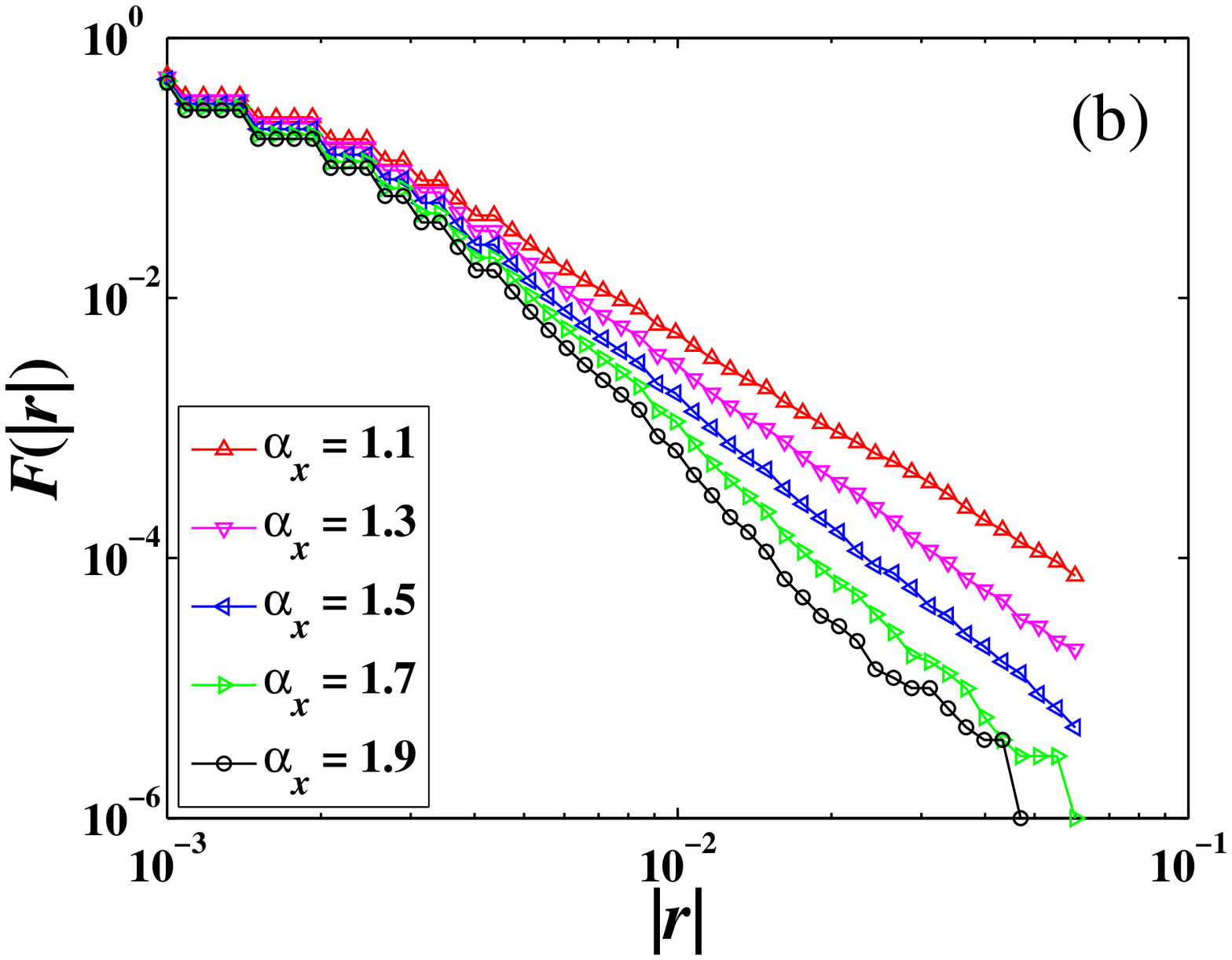}
\caption{\label{Fig:CDF:qG:G:DE} (Color online) Resultant
distribution of the trade-by-trade absolute returns $|r|$ when the
left part of $f(x)$ has a power-law tail, that is,
$f_L(x)=f_{qG}(x|\alpha_x,L)$. The right part of $f(x)$ is
$f_R(x)=f_{DE}(x|\lambda)$ in (a) and $f_R(x)=f_{G}(x|\sigma)$ in
(b), where $\lambda$ and $\sigma$ are determined respectively
according to Eq.~(\ref{Eq:x0:qG:DE}) and Eq.~(\ref{Eq:x0:qG:G}) for
different values of $\alpha_x$.}
\end{figure}

In this case, the left part of $f(x)$ has a power-law tail with
$f_L(x)=f_{qG}(x|\alpha_x,L)$ and we have two combinations for
$f_L(x)$ and $f_R(x)$, which are $\{f_{qG}(x|\alpha_x,L),
f_{DE}(x|\lambda)\}$ and $\{f_{qG}(x|\alpha_x,L),
f_{G}(x|\sigma)\}$. The simulation procedure is the same as in
Section \ref{S2:MFM:Case1}. The resultant empirical distributions of
the trade-by-trade absolute returns $|r|$ for each combination are
depicted in Fig.~\ref{Fig:CDF:qG:G:DE}(a) and (b), respectively.
Nice power-law tails are observed in all the distributions. For each
combination, the tail exponent $\alpha_r$ increases with $\alpha_x$.
Comparing the distributions in the two plots, no significant
difference can be identified in the return distributions with the
same value of $\alpha_x$. In other words, the shape of $F(|r|)$ or
equivalently $f(r)$ is fully determined by
$f_R(x)=f_{qG}(x|\alpha_x,L)$. It is not out of expectation that
there is no scaling in the distributions of the normalized returns,
$F(|g|)$, and we thus do not show them here.

In summary, our simulations confirms that the return distribution is
mainly determined by the distribution of the relative prices of
incoming orders placed within the order book with $x\leqslant0$.
Heavier tail in $f_L(x)$ will result in heavier tails in $f(r)$.
Only when $f_L(x)$ has a power-law tail, $f(r)$ will have power-law
tails.

\section{The return distributions at different timescales}
\label{S1:MFM:PDFs}

We now investigate the return distributions at different timescales
for the MF model with the parameters being extracted from real data
\cite{Mike-Farmer-2008-JEDC}. Specifically, the values of model
parameters used in this section are the following: $\alpha_x=1.3$,
$\sigma_x=0.0024$, $H_s=0.8$, $A=1.12$, and $B=0.2$. We stress that
$x$ follows the Student distribution in the standard MF model. The
simulation procedure is the same as described in the previous
section. We consider the normalized returns rather than the returns
{\em{per se}} for convenience.

We adopt the mid-price of the best bid $\pi_b(t)$ and best ask
$\pi_a(t)$ as the logarithmic price at time $t$ after a transaction
occurs:
\begin{equation}
I(t) = \frac{\pi_b(t) + \pi_a(t)}{2}~,
 \label{Eq:St}
\end{equation}
where $t$ is the event time corresponding to single trades. The
event-time return after $\Delta{t}$ trades is then defined as the
logarithmic price change:
\begin{equation}
r_{\Delta{t}}(t) = I(t)-I(t - \Delta{t})~.
 \label{Eq:r}
\end{equation}
Here we deal with the standardized returns
\begin{equation}
g_{\Delta{t}}(t) =
 [{r_{\Delta{t}}(t)-\mu_{\Delta{t}}}]/{\sigma_{\Delta{t}}}~,
 \label{Eq:gi}
\end{equation}
where $\mu_{\Delta{t}}$ and $\sigma_{\Delta{t}}$ are respectively
the mean and the standard deviation of returns $r_{\Delta{t}}$. For
simplicity, we drop the subscript $r_{\Delta{t}}$ below.

\subsection{Probability distributions of trade-by-trade returns}
\label{S2:MFM:PDF1}

\begin{figure}[htb]
\centering
\includegraphics[width=6cm]{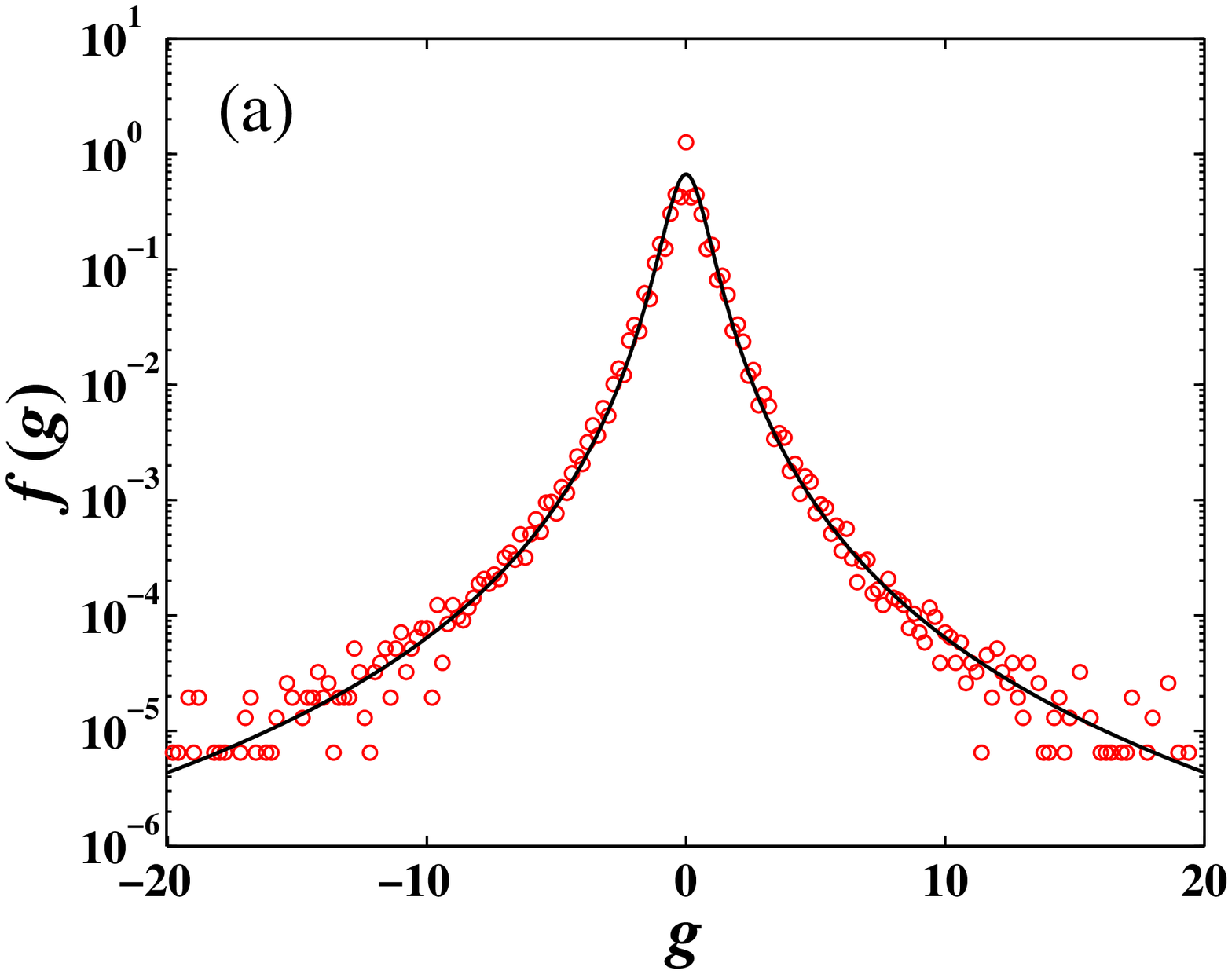}
\includegraphics[width=6cm]{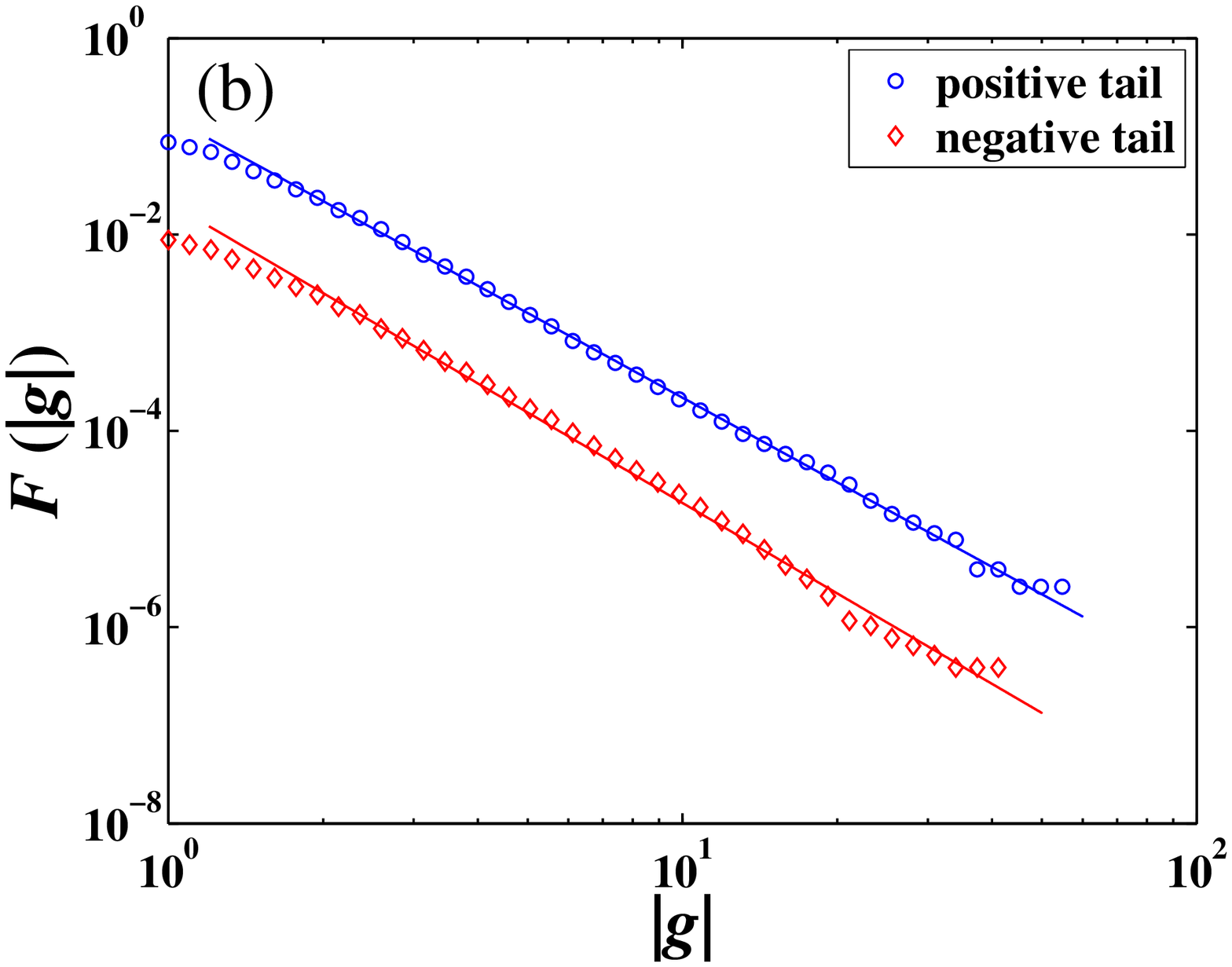}
\caption{\label{Fig:CDF:1} Empirical probability density function of
the normalized trade-by-trade returns. Panel (a): Empirical
probability density function $f(g)$ of the normalized returns $g$.
The solid line is the Student density with $\alpha_r = 2.9$ and $L =
3.3$. Panel (b): Empirical cumulative distributions $F(|g|)$ for
positive and negative normalized returns $g$. The solid lines are
the least squares fits of power laws to the data with $\alpha_r^+ =
2.87 \pm 0.02$ for the positive tail and $\alpha_r^- = 3.06 \pm
0.03$ for the negative tail. }
\end{figure}

We first focus on $\Delta{t}=1$. The empirical probability density
function $f(g)$ is estimated, as shown in Fig.~\ref{Fig:CDF:1}(a).
We find that $f(g)$ can be well modeled by a Student density.
Nonlinear least-squares regression gives $\alpha_r = 2.9$ and $L =
3.3$. The fitted curve is drawn on the left panel. According to
Fig.~\ref{Fig:CDF:1}(a), the Student density fits nicely the tails
of the empirical density $f(g)$. The fitted model deviates from the
empirical density remarkably for small values of $|g|$. If we
amplify the central part for small $|g|$ with finer binning, the
shape of $f(g)$ looks like a Mexican hat. This is the very character
of return distributions of individual stocks caused by the
discreteness of price changes in units of tick size. This intriguing
structure was reported for common stocks in the US market
\cite{Plerou-Gopikrishnan-Amaral-Meyer-Stanley-1999-PRE} and in the
Chinese market \cite{Gu-Chen-Zhou-2008b-PA}.

For large values of $|g|$, the Student density function $f(g)$
approaches power-law decay in the tails:
\begin{equation}
f(g) \sim \left\{
 \begin{array}{ccc}
 (-g)^{-(\alpha_r^- + 1)} && {\rm{for}}~~g<0\\
 (+g)^{-(\alpha_r^+ + 1)} && {\rm{for}}~~g>0
 \end{array}
 \right..
 \label{Eq:fg}
\end{equation}
The empirical cumulative distributions $F(|g|)$ of positive $g$ and
negative $g$ are illustrated in Fig.~\ref{Fig:CDF:1}(b). Both
positive and negative tails decay in a power-law form with
$\alpha_r^+ = 2.87 \pm 0.02$ and $\alpha_r^- = 3.06 \pm 0.03$, which
are in line with the tail exponent $\alpha_r$ estimated from the
Student model. These results indicate that the standardized returns
obey the (inverse) cubic law.

\subsection{Probability distributions of trade-aggregated returns}
\label{S2:MFM:PDF2}

We now turn to investigate the distributions of the standardized
trade-aggregated returns $g_{\Delta{t}}$, where $\Delta{t}$ spans
several trades. By varying the value of $\Delta{t}$, we are able to
compare the PDF's at different timescales. Specifically, we compare
the PDFs for $\Delta{t}= 2$, $4$, $8$ and $16$ trades with that for
$\Delta{t}=1$ trade. As listed in the second column of
Table~\ref{Tb:RNA}, the kurtosis of each PDF is significantly
greater than that of the Gaussian distribution whose kurtosis is 3,
indicating a much slower decay in the tails. In addition, the
kurtosis decreases with respect to the scale $\Delta{t}$. Very
similar leptokurtic behavior exists in real stock markets
\cite{Mantegna-Stanley-2000}.

\begin{table}[htp]
 \centering
 \caption{Characteristic parameters for trade-aggregated returns.}
 \medskip
 \label{Tb:RNA}
 \centering
 \begin{tabular}{cccccccccccc}
 \hline \hline
    \multirow{3}*[3.2mm]{$\Delta{t}$} & \multirow{3}*[3.2mm]{Kurtosis} &
    & \multicolumn{2}{@{\extracolsep\fill}c}{$q$-Gaussian}   &
    & \multicolumn{2}{@{\extracolsep\fill}c}{Positive tail}  &
    & \multicolumn{2}{@{\extracolsep\fill}c}{Negative tail}\\
    \cline{4-5} \cline{7-8} \cline{10-11}
  & && $L$ & $\alpha_r$ & &  Scaling range & $\alpha_r^+$ & &  Scaling range & $\alpha_r^-$
  \\\hline
  1  & 19.68 && 3.3 & 2.9 && $1.5 \leqslant{g}\leqslant 50.1$ &  $2.87 \pm 0.02$ && $1.5 \leqslant{-g}\leqslant 39.8$ & $3.06 \pm 0.03$\\
  2  & 19.52 && 3.1 & 3.0 && $1.5 \leqslant{g}\leqslant 36.3$ &  $2.90 \pm 0.02$ && $1.5 \leqslant{-g}\leqslant 30.2$ & $3.15 \pm 0.04$\\
  4  & 17.06 && 3.2 & 3.1 && $1.5 \leqslant{g}\leqslant 27.7$ &  $2.94 \pm 0.03$ && $1.7 \leqslant{-g}\leqslant 22.9$ & $3.24 \pm 0.04$\\
  8  & 13.91 && 2.9 & 3.2 && $1.7 \leqslant{g}\leqslant 15.8$ &  $3.07 \pm 0.04$ && $1.7 \leqslant{-g}\leqslant 15.9$ & $3.42 \pm 0.05$\\
  16 & 10.72 && 2.5 & 3.5 && $1.7 \leqslant{g}\leqslant 12.1$ &  $3.39 \pm 0.05$ && $1.9 \leqslant{-g}\leqslant  7.6$ & $3.77 \pm 0.07$\\
  \hline\hline
 \end{tabular}
\end{table}

\begin{figure}[htb]
\centering
\includegraphics[width=6cm]{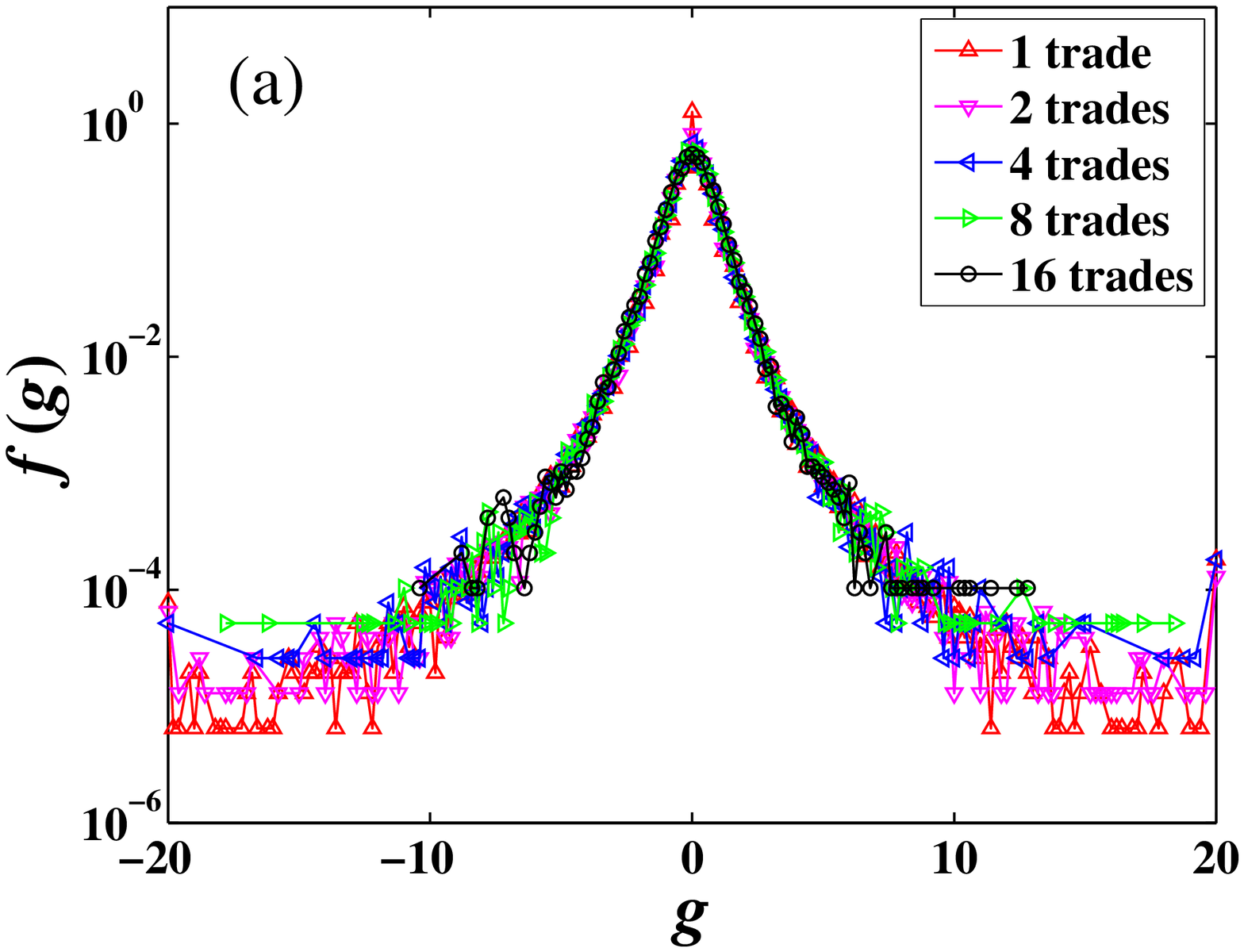}
\includegraphics[width=6cm]{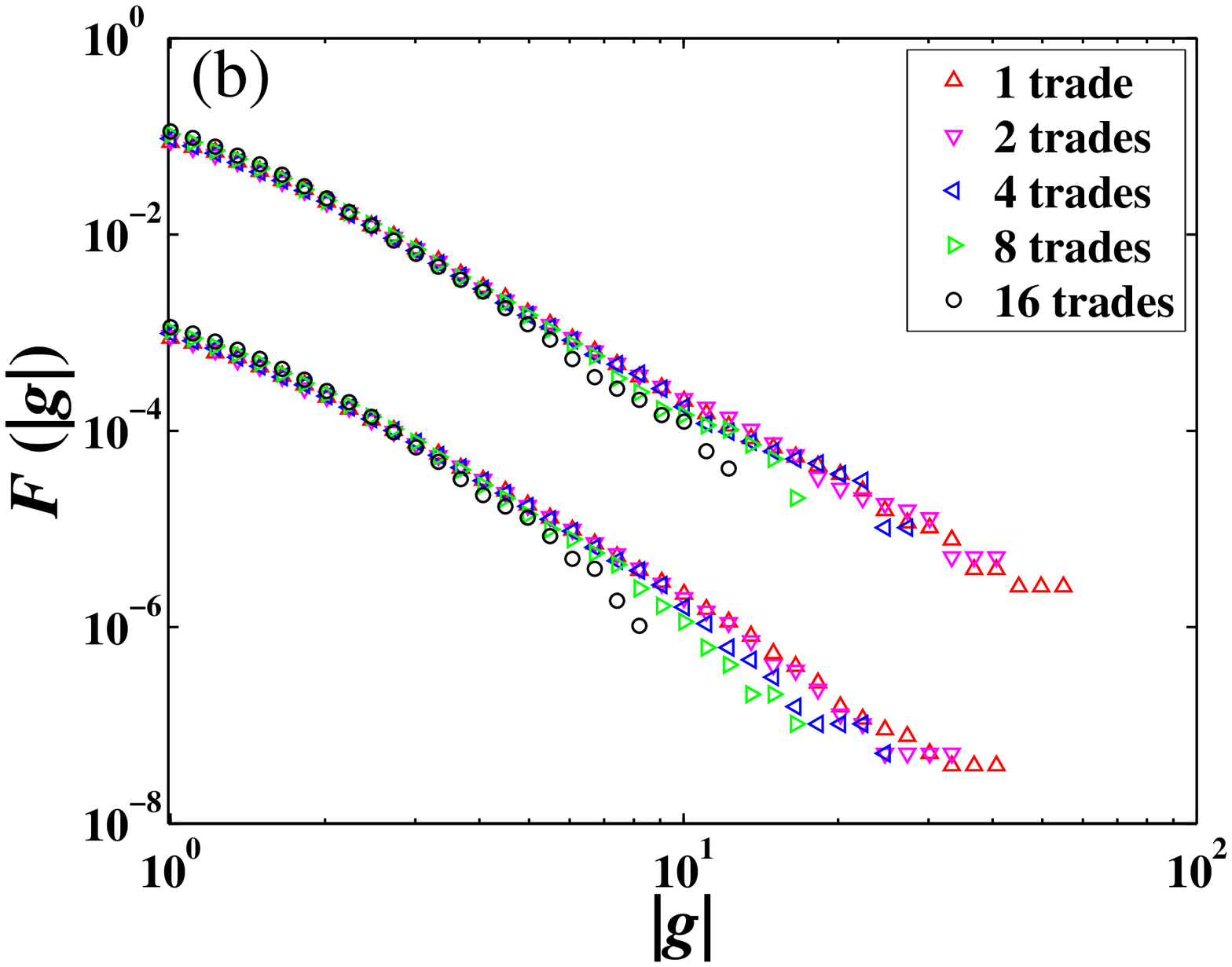}
\caption{\label{Fig:CDF:20} (Color online) Empirical distributions
of aggregated returns at different time scales $\Delta{t} = 1$, $2$,
$4$, $8$ and $16$. Panel (a): Empirical densities $f(g)$ of the
aggregated returns. Panel (b): Empirical cumulative distributions
$F(|g|)$ for positive (upper cluster of curves) and negative (lower
cluster of curves) returns $g$.}
\end{figure}

The empirical $f(g)$ functions for different time scales $\Delta{t}$
are illustrated in Fig.~\ref{Fig:CDF:20}(a). We represent the
distribution of one-trade returns for comparison. It is evident that
the tail is heavier with the decrease of $\Delta{t}$. This
phenomenon can also be characterized by the kurtosis of the
distributions. We also notice that the PDF for $\Delta{t}=16$ decays
slower than exponential. We have fitted the five curves using the
Student density model (\ref{Eq:Student}) and the estimated
parameters $L$ and $\alpha_r$ are listed in Table~\ref{Tb:RNA}. In
Fig.~\ref{Fig:CDF:20}(b), we study the tail distributions of the
normalized returns $g$. It is observed that both positive and
negative tails decay in power-law forms. We have estimated the tail
exponents, which are also presented in Table~\ref{Tb:RNA}. Note that
the scaling range decreases with increasing $\Delta{t}$, which is
also observed for two Korean indexes \cite{Lee-Lee-2004-JKPS} and 23
individual Chinese stocks \cite{Gu-Chen-Zhou-2008b-PA}. As expected,
the tails that are characterized by their tail exponents $\alpha_r$
decay faster for larger $\Delta{t}$, which is consistent with the
behavior of kurtosis.

The results obtained so far show that, the mock stock simulated from
the MF model shares striking similarity in the price dynamics with
the Chinese stocks \cite{Gu-Chen-Zhou-2008b-PA}. However, there are
also minor discrepancies. First, the tail exponents of the mock
stock are less than that of the Chinese stocks with about 0.2.
Second, the return distributions are slightly left-skewed while that
of the Chinese stocks are slightly right-skewed. It is not clear if
these discrepancies stem from the simplicity of the MF model or are
just a reflection of the fact that the parameters of the MF model
are not extracted from the Chinese stocks. The analysis presented
here shows that the MF model is very powerful and universal.

\section{Conclusion}
\label{S1:MFM:Conc}

In conclusion, we have studied the return distributions of mock
stocks in the Mike-Farmer model based on extensive simulations. We
found that the power-law tails of the return distribution $f(r)$ in
the MF model are caused by the power-law tail in the left part
$f_L(x)$ of the distribution of the relative prices of incoming
orders, no matter the right part $f_R(x)$ has a power-law tail or
not. In addition, we found that the distributions of returns in the
MF model at different timescales in units of trades can be modeled
by Student distributions, whose tail exponents are close to the
well-known cubic law and increase with timescale. The behavior of
return distributions is comparable to that of the real data.

\bigskip
{\textbf{Acknowledgments:}}

This work was partly supported by the National Natural Science
Foundation of China (Grant No. 70501011), the Fok Ying Tong
Education Foundation (Grant No. 101086), and the Program for New
Century Excellent Talents in University (Grant No. NCET-07-0288).

\bibliography{E:/Papers/Auxiliary/Bibliography}

\end{document}